\newif\ifblind
\blindfalse 

\ifblind
\def\wadi{WDS\xspace}
\def\wadilong{Water Distribution System\xspace}
\def\sutd{Uni X\xspace}
\def\sutdlong{University X\xspace}
\def\singapore{Country X\xspace}

\else
\def\wadi{WaDi\xspace}
\def\wadilong{Water Distribution\xspace}
\def\sutd{SUTD\xspace}
\def\sutdlong{Singapore University of Technology and Design\xspace}
\def\singapore{Singapore\xspace}

\fi

\documentclass[sigconf]{acmart}
\setcopyright{none}
\renewcommand\footnotetextcopyrightpermission[1]{} 

\usepackage{multicol}
\usepackage{amsmath}
\usepackage{mathtools}
\usepackage{commath}
\usepackage{balance}
\usepackage{wasysym}
\usepackage{booktabs}
\usepackage{color}
\usepackage{stackrel}
\usepackage{siunitx}
\usepackage{xspace}
\usepackage{tabulary}
\usepackage{url}
\usepackage{graphicx}
\usepackage{adjustbox}
\usepackage[inline]{enumitem}
\usepackage{subcaption}
\usepackage[shortcuts]{extdash}  

\newcolumntype{R}[2]{%
    >{\adjustbox{angle=#1,lap=\width-(#2)}\bgroup}%
    l%
    <{\egroup}%
}

\newcommand{\Paragraph}[1]{\smallskip\noindent{\bf #1.}}

\newcommand{\ie}{i.\@\,e.,\@\xspace}
\newcommand{\eg}{e.\@\,g.,\@\xspace}

\def\hb{\hbox to 10.7 cm{}}

\newcommand{\tcircle}[1]{\raisebox{.5pt}{\textcolor{SUTDred}{\textcircled{\raisebox{-.9pt} {#1}}}}}

\newcommand{\Botnet}{CPSBot\@\xspace}
\newcommand{\Botnets}{CPSBots\@\xspace}

\newcommand{\bn}{cpsb\@\xspace}
\newcommand{\CC}{C\&C\@\xspace}
\newcommand{\CCs}{C\&Cs\@\xspace}

\usepackage{pifont}
\newcommand{\cmark}{\ding{51}}%
\newcommand{\xmark}{\ding{55}}%

\definecolor{SUTDred}{RGB}{153,0,51}

\begin{document}

\title{ DeI: A Framework for Coordinated Cyber-Physical Attacks
\\on Distributed Industrial Control Systems}
\title{Publish-Subscriber Botnets for Real-Time Control Attacks}
\title{Taking Control: Design and Implementation of Closed-Loop-Control Botnets for CPS}
\title{Taking Control: Design and Implementation of Botnets\\ for Industrial
IoT}
\title{Taking Control: Design and Implementation of Botnets\\ for
Cyber-Physical Attacks with \Botnet}


\author{Daniele Antonioli}
\orcid{0000-0002-9342-3920}
\affiliation{%
  \institution{Singapore University of Technology and Design}
}
\email{daniele_antonioli@mymail.sutd.edu.sg}

\author{Giuseppe Bernieri}
\orcid{0000-0003-4801-0913}
\affiliation{%
  \institution{Roma Tre University}
}
\email{giuseppe.bernieri@uniroma3.it}

\author{Nils Ole Tippenhauer}
\orcid{0000-0001-8424-2602}
\affiliation{%
  \institution{Singapore University of Technology and Design}
}
\email{nils_tippenhauer@sutd.edu.sg}

\renewcommand{\shortauthors}{D. Antonioli et al.}

\begin{abstract}

  Recently, botnets such as Mirai and Persirai targeted IoT devices on a large scale.
  We consider attacks by botnets on cyber-physical systems (CPS),
  which require advanced capabilities such as controlling the physical
  processes in real-time. Traditional botnets are not suitable for
  this goal mainly because they lack process control capabilities, are
  not optimized for low latency communication, and bots generally do
  not leverage local resources. We argue that such attacks would
  require \emph{cyber-physical} botnets. A cyber-physical botnet needs
  coordinated and heterogeneous bots, capable of performing
  adversarial control strategies while subject to the constraints of
  the target CPS.

  In this work, we present \Botnet, a framework to build cyber-physical botnets.
  We present an example of a centralized \Botnet targeting a centrally
  controlled system and a decentralized \Botnet targeting a system
  distributed control.  We implemented the former \Botnet using MQTT
  for the \CC channel and Modbus/TCP as the target network protocol
  and we used it to launch several attacks on real and simulated
  \wadilong.  We evaluate our implementation with \emph{distributed
    reply} and \emph{distributed impersonation} attacks on a CPS, and
  show that malicious control with negligible latency is possible.

\end{abstract}


\maketitle

\section{Introduction}
\label{sec:intro}

Botnets are still one of the major threats in the cyber-security landscape.
IT botnets take advantage of Internet services such as IRC, HTTP, email,
and DNS to achieve different goals including information and identity
theft, spam, DDoS and malware distribution~\cite{silva2013botnets}.
The Mirai and Persirai IoT botnets are typical examples of IT botnets
for DDoS~\cite{bertino2017botnets}. In the OT space, we have seen
advanced malware such as Stuxnet~\cite{falliere2011stuxnet} and
Blackenergy~\cite{lee2016analysis,nazario2007blackenergy} using botnet\-/like
components to affect the availability of the target cyber-physical system (\eg
DoS and damaging the equipment).

However, we have not seen botnets capable of cyber-physical system
(CPS) \emph{adversarial control}. We claim that current botnet designs
are not sufficient to achieve this goal, mainly because high-impact
attacks on a CPS require different strategies than conventional
cyber-security attacks~\cite{slay2007lessons,gollmann2015cyber}. These
strategies translate into additional requirements that are not
addressed by conventional botnet designs. For example, a conventional
botnet does not differentiate its bots according to the capabilities
of the infected devices and does not allow coordinated interactions
among the bots. We think that it is beneficial to introduce a new
class of botnets, defined as \emph{cyber-physical} botnets, designed to
overcome those additional challenges. We expect that better
understanding of capabilities and shortcomings of cyber-physical botnets will
raise awareness with stakeholders of threatened systems, and allow the
defenders to design more suitable countermeasures.

In this paper, we present \textbf{\Botnet }, a framework to build
cyber-physical botnets. \Botnet enables to build botnets with heterogeneous
and coordinated bots able to take over the control of a CPS. \Botnet
is generic over the target CPS, it allows to develop botnets with
different network architectures and to use different adversarial control
strategies. We underline two of the most important design choices that
we made to satisfy our requirements. Firstly, we use a novel command
and control channel based on the \emph{publisher\-/subscriber (PubSub)}
paradigm~\cite{birman1987exploiting,eugster2003many} to get
precise coordination among bots with minimal overheads. Secondly, we define
a set of orthogonal functionality that we call \emph{traits} to customize the
development of a \Botnet. This modular approach is used to exploit
the functionalities offered by different infected devices of a cyber-physical
system and to customize the \CC servers.

Our attacker model considers a botmaster that already managed to infect
the target devices (how is outside the scope of this work). We present
two design examples of \Botnets targeting a centrally controlled system
and a system with distributed control.
We compare our examples against the traditional counterparts
and we motivate why we think that the latter are not
sufficient to enable adversarial control of the target systems.
We provide an implementation of the centralized botnet where an attacker is
coordinating infected gateway devices to influence the distribution of water
across remote substations. We use MQTT for the \CC protocol and we target the
Modbus/TCP industrial protocol.

We used our implementation to perform two coordinated cyber-physical attacks on
real and simulated \wadilong. The first attack is defined as
\emph{distributed impersonation} and the attacker is able to simultaneously
impersonate geographically-sparse remote terminal units. The second attack is
defined as \emph{distributed replay} and the attacker is able to reply values
and control actions across (potentially heterogeneous) devices in different
substations.

We argue that traditional botnet metrics, such as number of bots and DoS
bandwidth, are not sufficient to evaluate a cyber-physical
botnet. Hence, we define our own set of quantitative metrics suitable
to evaluate the \Botnet framework such as adversarial control period and
additional delay introduced by \Botnet and we use those metrics to evaluate our
cyber-physical attacks.

We summarize our contributions as follows:

\begin{itemize}

\item  We propose \emph{\Botnet } a framework to design
  cyber-physical botnets. Our framework addresses the
        extra\-/requirements introduced when attacking a cyber-physical system
        such as adversarial control and bots coordination capabilities.
        We use a publisher-subscriber command and control channel and
        traits to address those extra-requirements.

\item We design a centralized and a decentralized \Botnets targeting a
    centrally controlled system and a system with distributed control. We implement
    the former botnet optimizing it for precise coordination among
    bots using MQTT features such as quality of
    service, persistent sessions and asynchronous communication.

  \item We launch two coordinated cyber\-/physical
    attacks: dis\-tri\-bu\-ted impersonation and distributed replay to assess
    our implementation.
    Both attacks were performed on simulated and real water distribution
    testbeds, with minor code modifications. We evaluate them with
    our quantitative metrics for cyber-physical botnets.

\end{itemize}

This work is organized as follows: in Section~\ref{sec:background}, we
provide the background about botnets, cyber-physical systems and our
target water distribution system. In Section~\ref{sec:design} we present the design
of \Botnet starting from our problem statement and threat model.
We focus on the PubSub \CC channel and the \Botnet traits. Then, we show two examples
of centralized and decentralized \Botnets . We conclude the
section with our set of quantitative metrics for cyber-physical botnets.
In Section~\ref{sec:implementation}, we present how we
implemented a centralized \Botnet using MQTT and Modbus/TCP to attack our
target water distribution system.
In Section~\ref{sec:attacks} we describe the \Botnet attack phases and we
present and evaluate the distributed impersonation and
distributed replay attacks using the implemented centralized botnet.
We discuss several \Botnet attack strategies and optimizations
in Section~\ref{sec:discussion}. Related works are summarized in
Section~\ref{sec:related}, and we conclude the paper in
Section~\ref{sec:conclusions}.


\section{Background}
\label{sec:background}

\subsection{Botnets}
\label{sec:botnets}

A botnet is a network of compromised hosts (bots) that are managed by one
or more command and control (\CC) servers. The attacker (botmaster) is
connected to the \CC infrastructure and she is sending directives to the bots
through it. The channel of communication between the attacker and the bots
is called \emph{\CC channel}, and it is one of the most important parts of a
botnet~\cite{silva2013botnets}. A canonical way to classify a botnet is by its
network architecture. Most commonly a botnet is either \emph{centralized} or
\emph{decentralized}\footnote{In this context decentralized in synonymous of
distributed.}.

A centralized architecture has one \CC server that communicates with all
the bots. A client-server protocol is used for the \CC channel (\eg IRC
or HTTP). The main advantages of this setup are its low latency and ease
of coordination. The main weaknesses of this setup are its vulnerability
to single point of failure and network scalability issues when the number
of bots increases. Alternatively, in a decentralized architecture, all
compromised devices are used both as bots and \CC servers. The \CC channel
uses a peer-to-peer protocol (P2P) such that the bots establish an overlay
network. This architecture is self-scalable and does not suffer from single
point of failure. However, it might be difficult to implement (\eg hosts
behind NAT) and coordinate (\eg orders from multiple \CC). There are also
\emph{hybrid} architectures that provide a tradeoff between centralized and
decentralized schemes, and \emph{random} architectures where the bots are
not contacting the \CC server but they are waiting to be contacted by the
botmaster.

It is possible to represent the state of a botnet using a five-phases
\emph{lifecycle}. We have an \emph{initial infection}
phase, where the attacker exploits one or more vulnerabilities on a remote
machine. The remote machine becomes a bot candidate. In the \emph{second
infection} phase, the same infected machine is instructed to download and
execute different types of malware. If the malicious code is effective
then the infected machine becomes part of the botnet. In the third phase,
the bot contacts the \CC server and this process is defined as
\emph{rallying}. The rallying phase might be accomplished using static
addresses or dynamic addressing techniques such as DNS fast-flux, and domain
generation algorithms (DGA). The fourth
phase is the \emph{attack} phase, where the bot perform malicious activities
and might still exchange information with the \CC (\eg exfiltrate data). The
last phase is the \emph{maintenance} phase, where a botmaster might
modify the bot network configuration, upload new attack payloads and update
the cryptographic keys of the botnet.

The \CC channel uses either the same protocol of the target system or a custom
protocol (defined as neoteric~\cite{vormayr2017botnet}). The former
approach is, in general, less easy to detect because the malicious network
traffic is similar to the expected one. However, once the extra-traffic is
detected then the defender can isolate the offending traffic and try to
understand what is going on. The latter approach generates network traffic
that might stand out compared to the normal one. On the other hand, it is more
difficult for the defender to decode the information carried in a neoteric
packet~\cite{khattak2014taxonomy}.



\subsection{Cyber-Physical Systems (Security)}
\label{sec:cps}

Cyber-Physical Systems (CPS) are composed of heterogeneous devices that
are interacting with a physical process. These devices are typically
interconnected and they are programmed to perform general-purpose or
domain-specific tasks, including sensing, actuating, and networking. It is
possible to divide cyber-physical systems (CPS) in two categories: CPS with
a \emph{central} or \emph{distributed} control. In the first case, the CPS
uses a central monitoring and control infrastructure. For example, a water
distribution industrial control system is centrally controlled by a SCADA
server. In the second case, the CPS uses multiple controllers, each controller
manages a sub-system and it is able to communicate with the other controllers.
For example, a building automation system (BAS) is an example of a distributed
control system. There are also examples of distributed control system where
the control logic of one controller depends on signals coming from other ones.
These signals are called interlocks. A water treatment system is an example of
an interlock-based distributed control system.

The usage of Internet-friendly protocols and commodity hardware
vastly increased the attack surface of cyber-physical systems
~\cite{luallen2013sans,leverett2011ics,neumann2007communication,byres2004myths}.
CPS are vulnerable to classic information
security threats, attacks targeting the underlying physical processes, and the
intersection between the two~\cite{humayed2017cyber}. In this paper, we are
interested in the latter type of attacks, defined as \emph{cyber-physical}
attacks (\eg attack over the network that permanently damages a local component).

\subsection{The \wadilong (\wadi)}
\label{sec:wadi}

The \wadilong (\wadi) testbed is a water distribution autonomous
systems built at \sutdlong (\sutd) in 2016. It is designed as a
down-scaled version of the water transmission and distribution system
operated in \singapore.  \wadi enables simulating different water
demand patterns, water hammer effects, changes in pipe pressures and
pipe leakages. Furthermore, it allows simulating water
pollution by means of organic and inorganic contaminants that can be
added to the distribution of water to the consumers.

\wadi is composed of three sub-processes: \emph{water supply},
\emph{water distribution}, and \emph{water return}. The first stage operates
by taking the source water from two elevated raw water tanks ($2500 L$ each) and
transferring it to two elevated reservoirs ($1250 L$ each). In this stage,
water quality analyzers are used to verify the incoming water quality. In
the second stage, the potable water is distributed
to six consumers tanks ($500 L$ each). The water demand of Each consumer tank
can be set independently and changed in real-time. In the third stage, the
water is collected in the single water return tank ($2000 L$) and then
optionally returned back to the supply stage.


\wadi has different types of sensors (\eg water level sensors, flow meters,
water quality sensors) and actuators (\eg water pumps, valves). The remote
terminal units are SCADAPack 334E devices (Schneider Electric), the gateway
devices are MOXA oncell G3111-HSPA and the industrial switches are MOXA ED5
205A. The \wadi supervisory network uses the Modbus/TCP industrial protocol.

\section{\Botnets for Cyber-Physical Attacks}
\label{sec:design}

In this section, we present our problem statement and the related
system and attacker model. Then, we focus on the \Botnet
\CC channel and on the traits. We show the high-level
architecture of two \Botnets: a centralized botnet for ICS and a decentralized
botnet for IoT. We conclude the section defining a set of quantitative metrics
derived from the design of our botnets.

\subsection{Problem Statement}
\label{sec:problem}


Our main challenge is related to
\emph{adversarial control} of a system that has a
physical process and several interconnected heterogeneous devices (\ie a
cyber-physical system). With adversarial control, we refer to the
attacker's capability to steer the physical process of the target
system into states of her choice. Typically, this requires both
understanding of the physical process, and a suitable control strategy
(e.g., using closed loop control). 

We think that traditional botnets are not sufficient to launch coordinated
attacks on CPS systems for several reasons. Firstly, they do
not address the problem of adversarial control, and they are instead
focusing mostly on system disruption (\eg DDoS) and observation (\eg
eavesdropping). Adversarial control of the system would require near
real-time exchange of eavesdropped data and malicious control commands
between distributed bots together with one or more adversarial controllers.
We argue that traditional botnets are not designed to provide such features.
Secondly, IT botnets consider their bots as a homogeneous set of devices
(\eg zombies), without exploiting their different hardware and software
capabilities~\cite{silva2013botnets}. However, in a cyber-physical botnet, some
bots have different roles such as influencing the physical process and spoofing
the monitoring system. Thirdly, traditional botnet designs are not optimized
for CPS constraints such as latency, packet size and throughput. In our case,
these constraints have to be taken into account and measured in some way.




In summary, we would like to build a new class of botnets able to
perform coordinated cyber-physical attacks. We define those botnets as
\emph{cyber-physical}. In general, a cyber-physical botnet have to resemble a
control system, and the main difficulties to be addressed in its design are
the following:

\begin{itemize}
    \item Introduce an adversarial and coordinated
        control strategy on a CPS in real-time.
    \item Exploit the diversity of the bots, including their specific hardware
        and software capabilities.
    \item Evaluate the tradeoff between richness of bots functionality and
        associated overheads using sound metrics.
\end{itemize}


\subsection{System and Attacker Models}
\label{sec:models}

Our system model focuses on two types of cyber-physical systems: a system with
centralized control and a system with distributed control.


\Paragraph{Centralized Control}
We consider a (potentially geographically distributed) system composed
of $n$ substations that are centrally controlled (\eg a water
distribution system). Each substation is controlled by a remote
terminal unit (RTU) that is able to read sensor values and control
actuators. There are no intrusion detection systems deployed in the
substations.  A central SCADA server is periodically monitoring the
substations and it can send commands to the RTUs. A network-based intrusion
detection system (IDS) is monitoring the inbound and outbound traffic from the
SCADA network.

\Paragraph{Distributed Control}
We consider an IoT system with distributed control (\eg a building
automation system) that is deploying $m$ devices. Each device has a
specific functionality such as monitor and control the temperature of
a room (\eg HVAC), video surveillance (\eg CCTC), and lighting control
system (\eg LCS).

\begin{figure}[tb]
    \centering
    \includegraphics[width=0.7\linewidth]{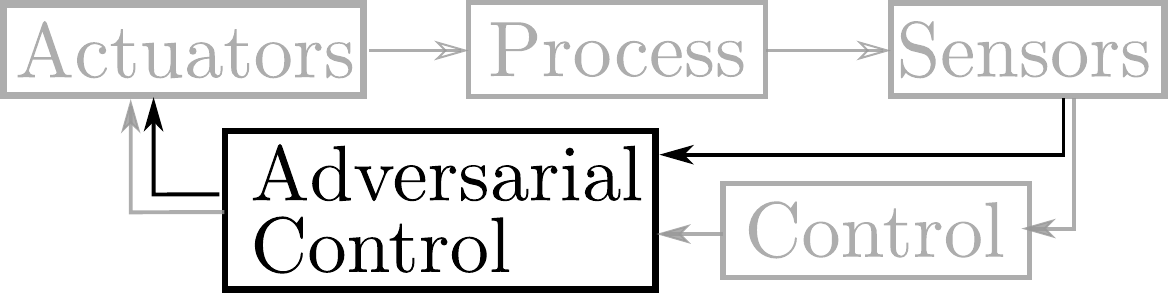}
    \caption{Attacker goal: real-time adversarial control strategy using
    a cyber-physical botnet.}
    \label{fig:attack-control}
\end{figure}

 \Paragraph{Attacker model} We consider an attacker who already completed
 the necessary steps to map the network of the target system, infect the
 devices, and perform the rallying phase (\eg the attacker is able to remotely
 contact the bots via the \CC). We believe that this is a reasonable attacker
 model to adopt both in the ICS and IoT scenario given the recent trends and
 surveys~\cite{sans2016survey,ics2016cert,sadeghi2015security,
 cui2010quantitative}. The main goal of the attacker is to use bots in a coordinated
 fashion to take control over the target cyber-physical system, adding
 an adversarial operation in the closed-loop control routines (see
 Figure~\ref{fig:attack-control}). In this setup the attacker, as the original
 controller, takes advantage of readings from multiple sensors (coming from
 different bots) and use this knowledge to send malicious actuation command to
 drive the system to an arbitrary state or sequence of states. Throughout the
 paper, we use the words attacker and botmaster interchangeably.


\subsection{\Botnet: PubSub \CC Channel}

We propose to use \emph{publish-subscribe (PubSub)} messaging pattern
for the \CC channel. In a PubSub scheme, there are three
entities: the publisher (sender), the subscriber (receiver) and the
broker (dispatcher). The communication is \emph{event-driven} and
there is a loose coupling between the sender and the receiver. This
scheme encourages the use of asynchronous communications in contrast
with client-server periodic request-response cycles
(\eg polling)~\cite{birman1987exploiting,eugster2003many}.
In our context, we consider a cyber-physical botnet with heterogeneous
bots subscribed to relevant events (\eg sensor values and actuator
states) and one or more \CC nodes publishing commands and updating
events based on those values in real-time. To the best of our
knowledge, the PubSub scheme has not been proposed for a botnet \CC
channel before, and is particularly well suited for our application.

Traditional IT botnets are using different \CC control protocols
mainly because they target different systems (\eg client-server
architecture). For example, IRC, DNS, email, and HTTP protocols are popular
choices~\cite{vormayr2017botnet}. In our system models (\eg ICS, and IoT)
those choices are either inapplicable (\eg the protocol is not spoken in the
system) or sub-optimal. As an illustration, HTTP is sub-optimal because it is a
client-server protocol (not event-driven) and it is not designed for
machine-to-machine communications (packet size is not a problem)\footnote{
We can apply a similar reasoning for the other traditional \CC protocols.}.

We list some crucial advantages that we think we would gain from
a PubSub control channel compared to conventional ones in the context of ICS
and IoT systems:

\begin{itemize}
    \item Flexible coordination among bots and \CC\@:
    \begin{itemize}
        \item Enabled by event-driven messaging.
        \item It allows using synchronous and asynchronous messaging schemes,
            multicast traffic, proactive and reactive bots coordination.
    \end{itemize}
    \item Compatibility with different botnet architectures:
    \begin{itemize}
        \item Enabled by loose coupling of publishers and subscribers.
        \item It allows building conventional (centralized, decentralized) and
            non-conventional (hybrid, random) botnets.
    \end{itemize}
    \item Addresses traditional and CPS botnets constraints:
    \begin{itemize}
        \item Enabled by the nature of PubSub intended for
            reliable and secure machine-to-machine
            communication~\cite{carzaniga2001design,pietzuch2002hermes}.
        \item It allows scaling the botnet size maintaining
            low computational and traffic overheads. For example, we can use
            anonymity~\cite{daubert2016anonpubsub},
            confidentiality and integrity mechanisms such as TLS
            or alternatives~\cite{ion2010supporting}.
    \end{itemize}
\end{itemize}


\subsection{\Botnet: Bot Traits}
\label{sec:architecture}

We already introduced the problem of cyber-physical bot heterogeneity that
translates into the need of bots supporting different functionalities. Those
functionalities derive from the role of a bot in an attack and they are
limited from the software and hardware capabilities of the bot.
We address the heterogeneity challenge using \Botnet
\emph{traits}. A trait represents a set of functionalities that
a \Botnet device might support. This enables to design a cyber-physical botnet
that is modular (\eg reuse same functionality across different devices),
extensible (\eg improve a functionality without affecting the others), and
composable (\eg mix functionalities in a single device). We note that traits
allow to customize \emph{both} the bots and the \CCs. We borrow the concept
of trait from object-oriented programming theory~\cite{scharli2003traits}.
We describe six traits that are relevant to our paper:



\Paragraph{Infiltrator} The Infiltrator affects the network configuration of
the infected device. For example, it might configure the bot as a malicious
proxy able to passively observe traffic, actively send payloads, forward traffic to
another network interface, and disconnect the bot from an arbitrary network.
In a typical setup, the number of Infiltrators scales linearly with the number of
\Botnet bots.

\Paragraph{Forger} The Forger tampers with the data coming from sensors,
actuators, and other connected devices. For example, it might locally
modify an actuator value while spoofing a remote monitoring server.
We note that the Forger takes advantage of different
hardware capabilities of the infected bots (\eg influence the physical
process). In a typical setup, the number of
Forgers is proportional to the number of \Botnet bots.


\Paragraph{Controller} The Controller takes care of the adversarial
control of the target system. In general, the Controller takes input from the
cyber-physical system and optionally from other \Botnet bots, and predicts the
future input-output state. The prediction could be computed using different
orthogonal techniques such as machine-learning classification, real-time
simulation, and state estimation techniques (\eg Kalman filtering). This 
functionality is typically implemented by the \CC in a centralized \Botnet, and 
by the infected controller devices in a decentralized \Botnet. A discussion 
about different prediction strategies is presented in 
Section~\ref{sec:learning}.



\Paragraph{Broker} The Broker functionality is used to coordinate the \Botnet
network. It asynchronously and synchronously sends event information to
all the botnet nodes. For example, a Broker manages the communication between
two bots without having them know each other and even
if one of them is disconnected from the \Botnet network. Typically, this
functionality is implemented by the \CC in a centralized botnet and by multiple nodes
(\eg broker clustering) in a decentralized botnet. We note that, an architecture with
multiple Brokers tolerates single-point-of-failure in the \Botnet network.

\Paragraph{Pub} The Pub allows the \Botnet devices to send data over the
botnet network asynchronously. The granularity of the published content can be
set (\eg publish an aggregate of sensor values versus a single sensor value).
Furthermore, the Pub could set the quality of service of each published
(sent) value. For example, the botmaster can coordinate the botnet using an
event-based priority scheme dependent by the message type (topic-based) or the
message value (content-based). This functionality is typically carried out by
all \Botnet nodes.

\Paragraph{Sub} The Sub allows the \Botnet devices to receive data
over the \Botnet network. Each Sub might subscribe to any information exchanged
in the \Botnet network, and get it on-demand, without sending a request all the
times (event-driven). The subscription process is pre-configurable to avoid
re-subscriptions after successive disconnections. Additionally, each Sub can
create a session with the Broker to let it cache lost messages and retrieve
them among re-connection.
For example, the \CC node might subscribe to status-critical information to be
informed when any of the bots is disconnected from the \Botnet network and react
accordingly. Similarly to the Pub functionality, the Sub functionality is
typically implemented by all \Botnet nodes.

In summary, we think that designing a botnet using traits is an effective
way to address the diversity of devices found in cyber-physical systems. For
example, we can use traits to differentiate the implementations of bots for
network spoofing and bots for adversarial control (we will see two concrete
examples in Section~\ref{sec:examples}). As a side benefit, the usage of
traits lowers the development costs of our \Botnets and this is a key factor
for effective cyber-physical attacks~\cite{gollmann2015cyber}.


\subsection{Centralized and Decentralized \Botnets}
\label{sec:examples}

\begin{figure*}[tb]
    \centering
    \begin{subfigure}[b]{0.46\linewidth}
        \includegraphics[width=1.0\linewidth]{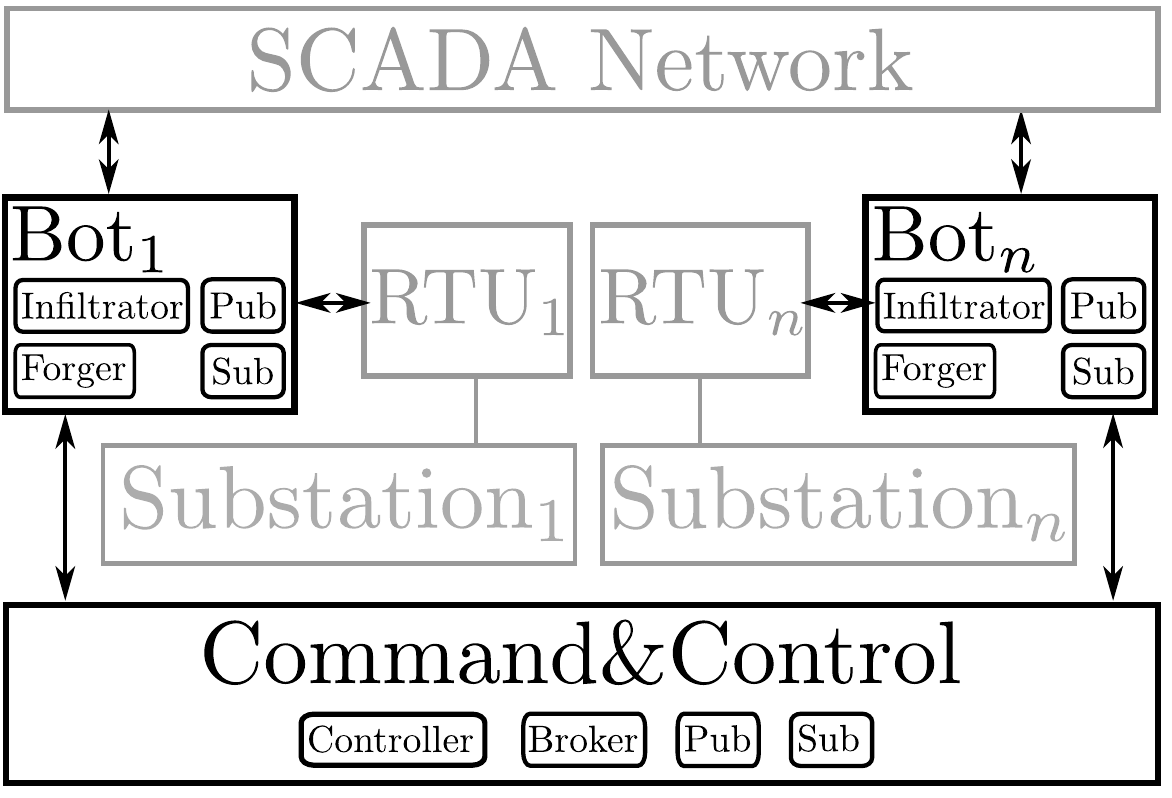}
        \caption{Centralized botnet targeting a distributed ICS (\eg water distribution system).}
        \label{fig:arch-centralized}
    \end{subfigure}
    \hfill
    \begin{subfigure}[b]{0.46\linewidth}
        \includegraphics[width=1.0\linewidth]{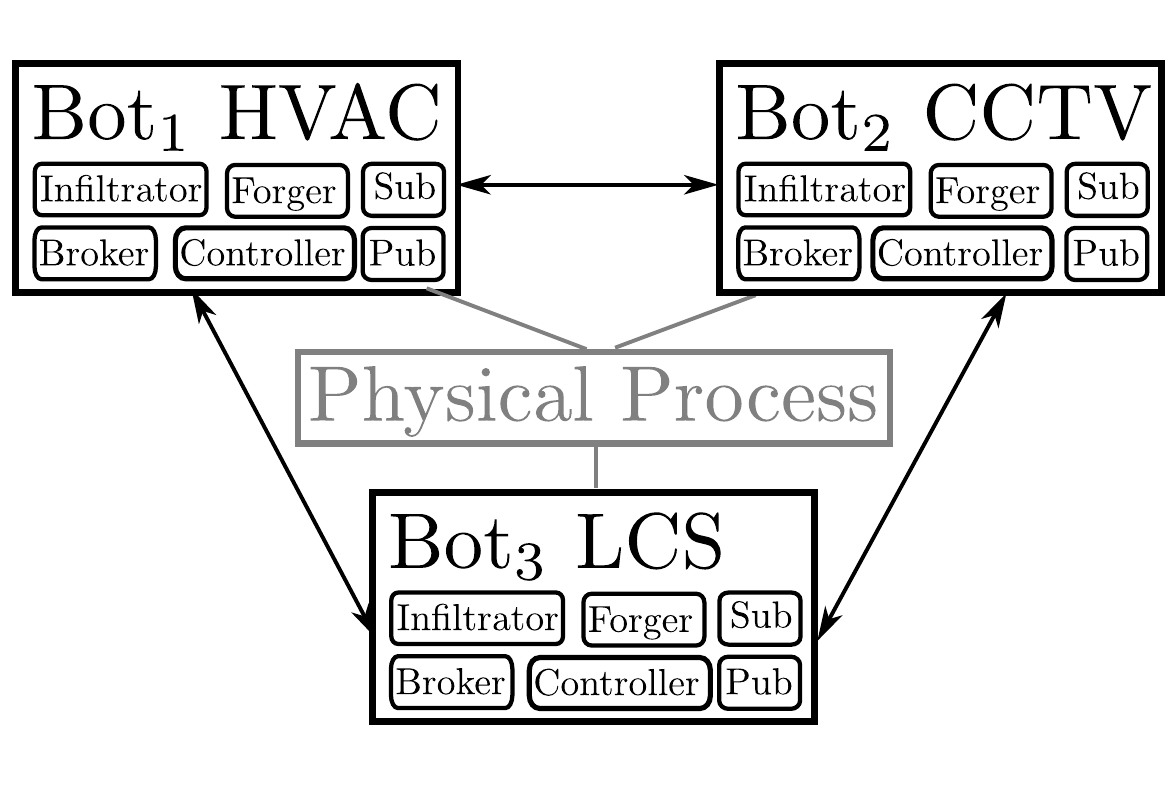}
        \caption{Decentralized botnet targeting an IoT system (\eg building automation        system)}
        \label{fig:arch-decentralized}
    \end{subfigure}

    \caption{High-level view of an ICS and IoT \Botnet. Grey boxes
    represent the targets. The black boxes are the devices controlled by
    the attacker.
    Rounded black boxes represent the traits supported by the
    \Botnet devices.}
    \label{fig:arch}
\end{figure*}

We now present two \Botnets: a centralized botnet attacking a water
distribution system (ICS) and a decentralized botnet attacking a building
automation system (IoT). We choose these examples because they share similar
security functions and weaknesses~\cite{brandstetter2017bas}. For the
sub-figures in Figure~\ref{fig:arch}, we represent the devices controlled by
the attacker with black squared boxes. The traits of the \Botnet devices are
represented as black boxes with round corners. The grey boxes represent the
targets. Both botnets derive from the system and attacker models presented in
Section~\ref{sec:models}.

\Paragraph{Centralized \Botnet} In this scenario, the attacker is using a
centralized architecture to attack a centralized control system (\eg water
distribution system). As depicted in Figure~\ref{fig:arch-centralized}, the
system is composed of $n$ remote substations and a central monitoring SCADA
network. Each substation has a remote terminal unit (RTU) that interacts with
sensors and actuators and a gateway device that connects the RTU to the
access router. The botmaster has compromised $n$ gateway devices (Bot$_1$,
\ldots , Bot$_n$). Each bot implements the Forger, Infiltrator, Pub and Sub
traits. The central \CC is managing all the bots and it is implementing the
Controller, Broker, Pub, and Sub traits. The bots are altering the state of
the physical process in real time with the help of the \CC while fooling
the central SCADA server that is periodically querying the RTUs. This
\Botnet design is different from a traditional one because each bot acts in
a different manner according to its substation. For example, the first bot
is mainly interested in the sensor and actuator values regarding the first
substation and it generates spoofed commands accordingly. In a traditional
setup, each bot will execute the same orders from the \CC regardless of which
substation it is affecting. More information about an implementation of
this botnet are presented in Section~\ref{sec:implementation} and the two
cyber-physical attacks are evaluated in Section~\ref{sec:attacks}.

\Paragraph{Decentralized \Botnet} In this case, the botmaster is using a
decentralized architecture to attack a distributed control system, in
particular, a building automation system (BAC). As we can see from
Figure~\ref{fig:arch-decentralized},
the BAC is composed of a heating, ventilation and air-conditioning
system (HVAC), an IP camera (CCTV) and a lighting control system (LCS). In
this case, each infected device acts both as a bot and as a \CC. All the bots
are implementing the Controller trait and they are locally computing their
adversarial estimations. Additionally, they implement the Infiltrator, Forger,
Sub and Pub traits. In this scenario, we take advantage of multiple brokers (\eg
every bot implements the Broker trait) to avoid single-point-of-failure. For
example, if one bot is compromised the others can still coordinate their
actions. This botnet is different from a traditional P2P botnet because each
bot has a specific control strategy depending on the infected device.
Additionally, each bot might share control information with the others
even if the target system is not interlock-based. For example, a botmaster
could use \emph{adversarial interlocks} to perform cascade cyber-physical
attacks (\eg induce an LCS blackout while tampering with the HVAC load).


\subsection{Our Quantitative Metrics}
\label{sec:metrics}

Conventional botnets are evaluated looking at factors such as number
and geo-locations of bots and \CC servers, malicious DNS activities, and malware
databases~\cite{nadji2013beheading,stone2011analysis}. However, we think that
to evaluate cyber-physical botnets we have to define additional
metrics. Table~\ref{tab:metrics} lists our set of quantitative metrics
that we are using in Section~\ref{sec:evaluation} to evaluate our
cyber-physical attacks. A checkmark (\cmark) in the CPS column indicates that the
metric is defined by us to address the \Botnet cyber-physical constraints.


\begin{table}[tb]
    \centering
    \caption{Quantitative metrics used to evaluate the \Botnet attacks. A checkmark (\cmark) in the CPS column indicates that the
metric is defined by us to address the \Botnet cyber-physical constraints.}
    \label{tab:metrics}
    \begin{tabulary}{0.46\textwidth}{rLc}

        \toprule
        \textbf{Symbol}   & \textbf{Metric description} & \textbf{CPS} \\
        \midrule
        $T_S$ [s]         & Period between two equal requests from the
        defender. & \cmark \\
        $T_C$ [s]         & Adversarial estimation period. & \cmark \\
        $T_M$ [s]         & Traffic manipulation period.& \cmark  \\
        $\Delta_{s}$ [ms] & Delay between the last valid packet and
        the first spoofed packet.& \cmark  \\
        $\Delta_{r}$ [ms] & Delay between the last valid request and
        the first spoofed response.& \cmark  \\
        $\mu_{d}$ [ms]    & Additional delay introduced by the \Botnet. & \cmark  \\
        $n_B$             & Number of bots. & \xmark\\
        $n_{e}$           & Number of IDS warnings and errors raised. & \xmark\\
        $\mu_{CPU}$       & Average CPU overhead for the bot. &\xmark\\
        $\mu_{RAM}$       & Average RAM overhead for the bot. &\xmark\\
        \bottomrule

    \end{tabulary}
\end{table}



\section{Design and implementation for ICS}
\label{sec:implementation}

In this section, we present an implementation of a centralized \Botnet
to attack a water distribution industrial control system.  In
particular, the system is controlled using the Modbus/TCP protocol. We start by
mapping the network architecture sketched in
Figure~\ref{fig:arch-centralized} to a water distribution
network. We then describe how we deal with the target industrial
protocol and the botnet \CC channel protocol. We conclude the section
showing how we implemented some specific PubSub features to better
coordinate the \Botnet bots.

\subsection{Network Topology}
\label{sec:dei-network}

\begin{figure*}[tb]
    \centering
    \includegraphics[width=0.6\linewidth]{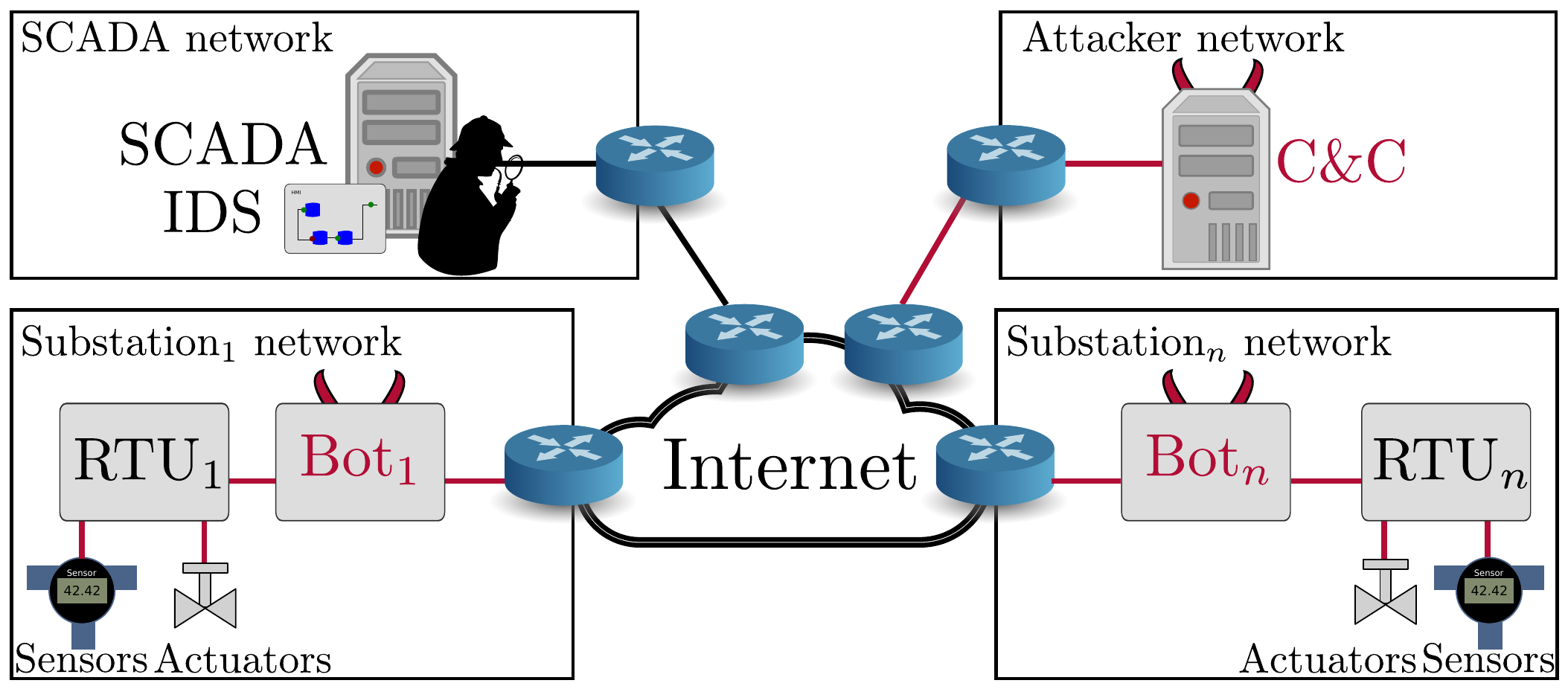}
    \caption{Centralized \Botnet attacking a water distribution ICS. There are $n$
    substations. Each substation has a remote terminal unit (RTU), and a
    compromised gateway device (Bot). The SCADA server is in the
    central control network. The attacker (via the \CC) coordinates the bots
    from a remote location using the red links.}
    \label{fig:ics-botnet}
\end{figure*}

Figure~\ref{fig:ics-botnet} shows the network topology of a water distribution
ICS already compromised by a centralized \Botnet. There are $n+2$ networks:
$n$ substation networks, the SCADA network, and the attacker network.
Each network has a border router connected to the Internet. The remote
terminal units (RTU) are managing local sensors and actuators and they are
communicating with the central SCADA server through gateway devices. A
network-based intrusion detection system (IDS) is monitoring the inbound and
outbound traffic from the SCADA network.
The attacker infected $n$ gateway devices that are sitting in between
the RTUs and the border routers. The botmaster uses the symbolic
links colored in red to communicate with the bots.

\subsection{Target Protocol: Modbus/TCP}
\label{sec:modbus}

We choose Modbus~\cite{modbus} and in particular Modbus/TCP as a target
industrial protocol because it is still widely used on actual industrial
plants\cite{modbus-usage}. Furthermore, Modbus/TCP is  adopted in \wadi,
the water distribution testbed that we use for our attacks. We understand
that Modbus (as many popular industrial protocols) is not secure by design.
However, we are not interested in discovering or underlying existent Modbus
vulnerabilities.


Modbus includes two data types: registers and bit fields.
Registers are 16-bit integers and they are either read-only (input registers)
or read and write (holding registers). Bit fields are either read-only
(discrete inputs) or read and write (coils). All data types are addressed like
an array in memory and the first array element is at offset zero. Modbus/TCP is
a client-server application layer protocol. A Modbus request has a Modbus/TCP
header that contains: the transaction number (set by the client and echoed
back by the server), the length of the payload, the protocol ID, and the slave
ID. The payload of the Modbus request addresses the requested data with a function
code, a memory offset, and a word count. A Modbus response contains a similar
Modbus/TCP header and its payload contains the same function code of the
corresponding request, a byte count, and the requested data.

Modbus operations are encoded with numeric function codes. The attack that
we present in Section~\ref{sec:attacks} focus on three of them: read coils
(\texttt{0x01}), read holding registers (\texttt{0x03}), and write single
coil ( \texttt{0x05}). Read coils is used by a Modbus client to read multiple
binary values from a Modbus server. Read holding registers are used by a Modbus
client to read multiple 16-bit registers from a Modbus server. Write single
coil is used by the Modbus client to write a single bit to a Modbus server.

It is worth mentioning that the Modbus packets are transmitted using network
ordering (\eg big endian) but the bit fields are stored in reverse bit order.
Hence, if the first byte of the coil memory contains the following bits
\texttt{01100011}, then the last (eighth) bit will map to the first coil and
its value will represent True (\texttt{1}), the seventh bit will map to the
second coil and its value will represent True (\texttt{1}), and so on.


\subsection{\CC Channel: MQTT}
\label{sec:dei-protocol}


We use the Message Queuing Telemetry Transport (MQTT) protocol for the
implementation of the \CC control channel. MQTT is a \emph{topic-based} PubSub
protocol, designed for low\-/bandwidth, high\-/latency Machine to Machine
(M2M) communication~\cite{mqtt}. By default, it runs over TCP, it
supports TLS and password-based authentication of clients.
All the messages exchanged in the \Botnet botnet are addressable
by topic, and their payload is data\-/agnostic. Topics can
be hierarchically organized with different paths and each path can contain
sub-paths. The set of all topics is called the topic tree, and its design is
key for an effective MQTT botnet coordination.

\begin{figure}[b]
    \centering
    \includegraphics[width=1.0\linewidth]{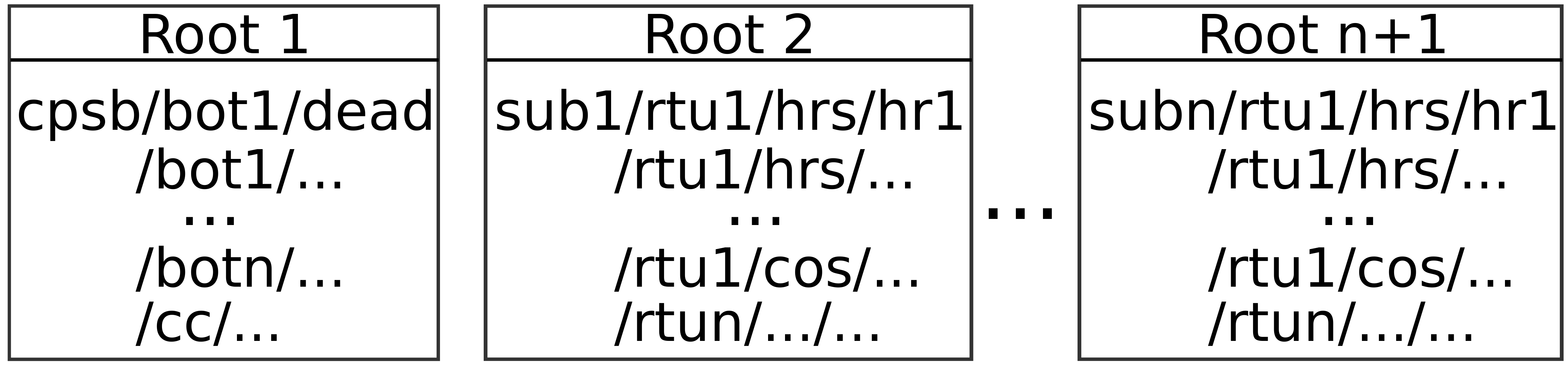}
    \caption{\Botnet hierarchical topics tree design with $n+1$
            paths. The \texttt{\bn} path takes care of the messages about
            the botnet devices. There are $n$ \texttt{sub} paths, each one
            takes care of the messages about a water distribution substation.}
    \label{fig:topics-tree}
\end{figure}

In Figure~\ref{fig:topics-tree} we present our implementation of a topic tree
suitable to manage our \Botnet attacking a water distribution system. It
includes $n+1$ paths, where $n$ is the number of
attacked substations. The \texttt{\bn} path contains one sub-path for each \Botnet
device (in this case $n+1$ sub-paths). For example, if a node subscribes to the
\texttt{\bn/bot1/dead} topic, then it will receive updates when the bot in
the first substation is disconnecting from the network (together with all the
other subscribers). Another usage of this path concerns the maintenance of the
bots. For example, we can use \texttt{\bn/bot1/sw} to send binary software
updates to the subscribed clients.
The \texttt{subx} paths contain the information
about the water distribution substations. Each sub-path manages a
device (only RTUs in this case) and each sub-sub-path manages sensor
and actuator values using the same
memory mapping of the target RTU. For example, the \texttt{sub1/rtu1/hrs} topics
contain all the messages regarding the values of the holding registers
of the first RTU.

We used \texttt{mosquitto} for the MQTT broker and \texttt{paho} for the MQTT
clients. Note that our MQTT setup does not depend on the target industrial
protocol and with minor modifications, it can be adopted for other physical
processes.

\subsection{Coordination of the \Botnet Nodes}
\label{sec:scripts}

MQTT provides several useful functionalities that we are using for coordinated
interactions among \Botnet bots. We comment five of them:

\Paragraph{Message QoS} Each publish and subscribe action can be configured
with a quality of service (QoS) value. There are three possible QoS values
for delivering a packet at most once (QoS = 0), at least once (QoS = 1), and
exactly once (QoS = 2). The default publish and subscribe QoS values are
0 and increasing QoS values result in a bigger protocol overhead.
Each client subscribes to topics with a QoS and receives published messages
with that QoS, even if the publisher is setting a higher QoS.
We use MQTT's QoS to build a priority scheme based on the message
topic. For example, we give maximum priority (QoS = 2) to messages about
bot disconnections and medium priority (QoS = 1) to messages about botnet
maintenance, critical sensors and actuators values, and low priority (QoS = 0)
to the other messages.

\Paragraph{Client sessions} MQTT consents to store in the broker(s) a
subscription session for each subscriber (client) using a unique ID.
We use this feature to optimize the clients' subscription process and
message recovery. If the client session is turned on (\eg by setting the
\texttt{clean\_session} flag to \texttt{False}) then the client subscribes
once to the topics of interest and it does not need to resubscribe upon
re-connection. Furthermore, the broker stores all the missed published
messages with QoS greater than 0 and it re-publishes them when the client
reconnects if the subscription was made with QoS greater than 0.

\Paragraph{Asynchronous connections}
MQTT supports both blocking (synchronous) and non-blocking (asynchronous)
connections to the broker. We decided to use non-blocking
clients and servers to optimize the usage of the \Botnets nodes. For example,
a bot might perform other tasks while waiting to establish a connection with a
broker.

\Paragraph{Subscription with wildcards}
MQTT supports meta-characters to efficiently subscribe
to multiple topics. For example, to subscribe to all coils messages from the
second substation a bot can subscribe to \texttt{sub2/+/cos/*}. The
\texttt{+} meta-character subscribes to all the topics at the current path level,
while the \texttt{*} meta-character subscribes to all the topics in the
current path level and below. We use this feature in combination with our topic tree
design to allow easy subscriptions to intra-substation and inter-substations topics.

\Paragraph{Parametric keep-alive interval}
MQTT is transported over TCP and sometimes one of the two hosts in a TCP
stream is not working properly. This situation is called half-open connection
problem and the MQTT keep-alive functionality is used to fix it.
Each client can communicate the (maximum) keep-alive time in seconds which
broker and client can endure without sending a message. The default keep-alive
time is 60 seconds, and if it is set to 0 then this mechanism is not used.
We use this feature to ensure that each communication link is working as
expected and to manage the relative geographic positions of the MQTT client and the
MQTT broker. For example, clients that are geographically closer to the broker may
be configured with a lower keep-alive value than the ones that are farther
apart.





\section{Case study: Attacks on ICS}
\label{sec:attacks}

In this section, we present two cyber-physical attacks performed on simulated
and real water distribution testbeds. The attacks use the centralized \Botnet
implemented in Section~\ref{sec:implementation}. We first performed simulated
attacks to speed-up the development time and to reduce the risk of damaging
actual components. Then we performed the same attacks on the \wadilong testbed
(presented in Section~\ref{sec:wadi}). We start this section by describing
the phases of \Botnet-based attack. Then we report on the initialization,
distributed impersonation and distributed replay phases. We conclude the
section with a quantitative evaluation of the presented attacks using the
metrics defined in Section~\ref{sec:metrics}.

\subsection{Phases of \Botnet-based Attack}
\label{sec:attack-example}

\begin{figure}[tb]
    \centering
    \includegraphics[width=1.0\linewidth]{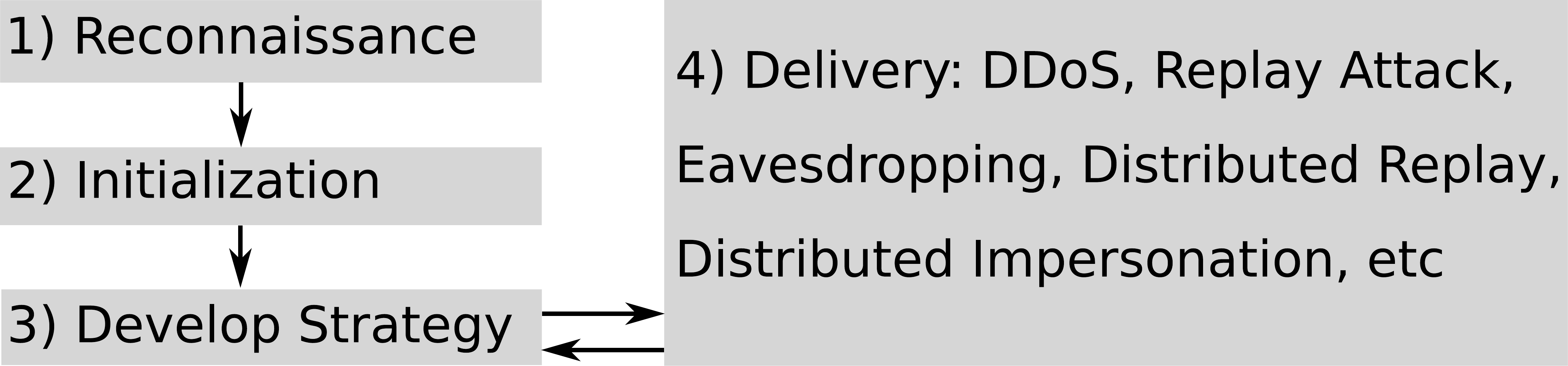}
    \caption{A \Botnet attack has four phases. Reconnaissance,
    Initialization, Develop Strategy, and Delivery. The first two phases are
    common to all \Botnet attacks. The last two are iterative.}
    \label{fig:dei-phases}
\end{figure}  

A \Botnet-based attack can be decomposed into four phases based on the
industrial control system cyber kill chain~\cite{sans2015icsckc}. The
attack phases are depicted in Figure~\ref{fig:dei-phases}). Firstly, we
have the \emph{Reconnaissance} phase. This is a preliminary phase where
the attacker tries to get as much information as possible about the target
system. Secondly, we have the \emph{Initialization} phase. In this phase, the
botmaster completes the initial infection, second infection, and rallying
phases. Thirdly, we have the \emph{Develop Strategy} phase. In this phase,
the bots observe the network and the physical process and develop different
attack options with the help of the \CC and the botmaster. Finally, we have
the \emph{Delivery} phase where the botmaster launches the attack(s) and tries
to reach one or multiple goals. In Figure~\ref{fig:dei-phases} we list several
traditional goals such as DDoS, replay and eavesdropping and cyber-physical
goals such as distributed replay and distributed impersonation. We note that
\Botnet allows delivering multiple non-interfering attacks at the same time (\eg
impersonate a device while eavesdropping the communications).

In the following section, we assume that the attacker already completed
the first two attack phases (that are common to all \Botnet attacks)
and we discuss two advanced cyber-physical goals: \emph{distributed
impersonation} and \emph{distributed replay}, Distributed impersonation
allows the attacker to impersonate multiple gateway devices at the same time
and coordinate their responses to the SCADA server.
(see Section~\ref{sec:impersonation}). Distributed replay enables the attacker to
programmatically replay messages across substations (see Section~\ref{sec:replay}).


\subsection{Initialization Phase}
\label{sec:init}

\begin{figure*}[tb]
    \centering
    \includegraphics[width=0.5\linewidth]{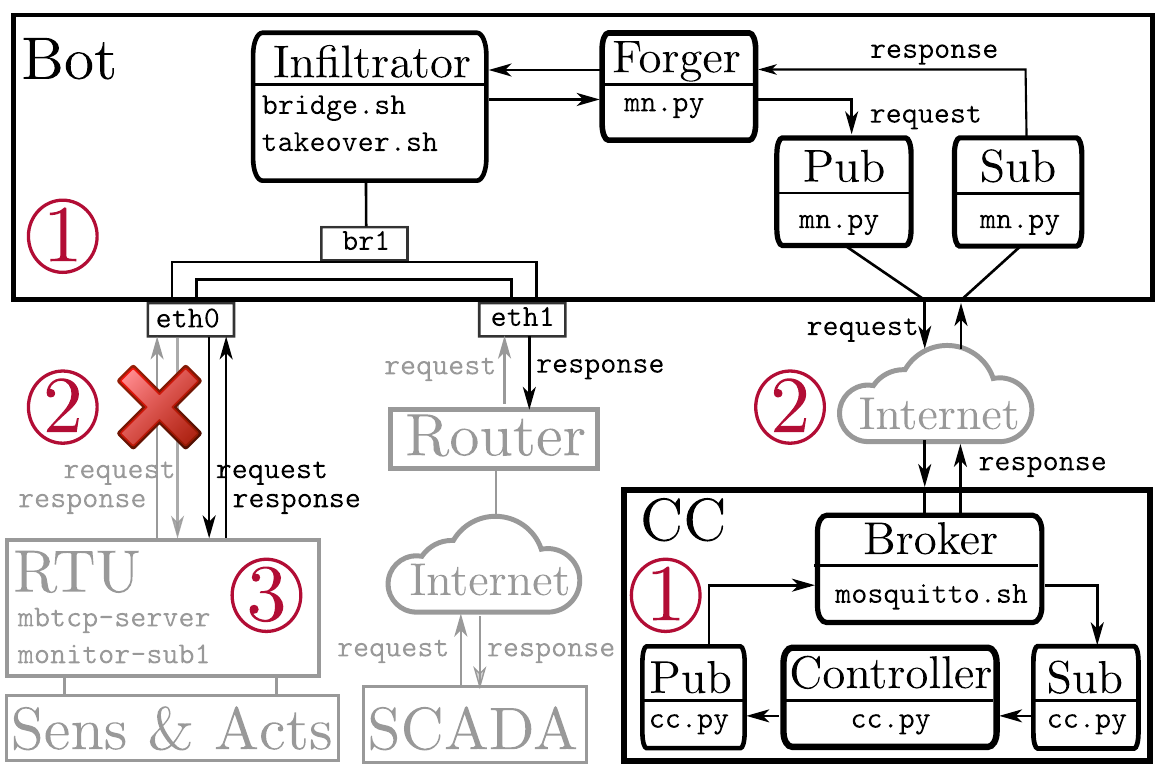}
    \caption{A \Botnet-based attack on the \wadilong.  Grey/black
      lines and boxes represent the ICS/\Botnet links and devices.  In the
      initialization phase, the bot is bridging between the RTU and
      the router, and the \CC starts the Broker process and the Pub
      and Sub clients  \tcircle{1}. In the distributed impersonation
      phase the bot disconnects the RTU from the network and start
      impersonating it by answering to the SCADA \texttt{requests}
      using the \texttt{responses} produced by the \CC Controller
      \tcircle{2}.  The same phases take place on the other substations at
      the same time. Once the botmaster impersonates all the RTUs then
      she can target the real ones \tcircle{3}.  }
    \label{fig:attack-sub1}
\end{figure*}

Figure~\ref{fig:attack-sub1} presents an attack scenario where a \Botnet bot
already infected a gateway device of \wadi. The bot,
as in Figure~\ref{fig:ics-botnet}, sits in between the remote terminal unit
(RTU) and the access router of the substation. The bot is implementing the
Infiltrator, Forger, Pub and Sub traits, indeed it is able to spoof the
packets coming from its network interfaces.

Part of the initialization phase is accomplished using a combination
of Ethernet bridging and forwarding techniques, see top-left \tcircle{1}
in Figure~\ref{fig:attack-sub1}.
Basically, an Ethernet bridge is a virtual switch that forwards all the
traffic from \texttt{eth0} to \texttt{eth1} and vice-versa. It can
be configured to act as a firewall at the link and network layers.
Once the bridge is established, then the bot is
able to observe all the traffic between the RTU and the access router through
the bridge interface (\texttt{br1}). The infected gateway devices are running
Linux and we used \texttt{bridge-utils}, \texttt{iptables},
and \texttt{ebtables} tools to configure the bridges on the bots. The
setup can be easily extended to bridge more than two interfaces.

Additionally, in order to optimize the \Botnet boot time, we are using the initialization
phase to start the functionalities of the command and control node.
Firstly, the \CC Broker is starting a \texttt{mosquitto} MQTT broker with a
customizable configuration file.
Then, the \CC Pub and Sub are starting asynchronous \texttt{paho}
MQTT clients to establish the \Botnet network. Finally, the \CC Controller starts
in idle-mode because there is no physical process data to be processed, see
bottom-right \tcircle{1} in Figure~\ref{fig:attack-sub1}.



\subsection{Distributed Impersonation Phase}
\label{sec:impersonation}

At the beginning of the distributed impersonation phase, the bot Infiltrator
isolates the RTU from the central SCADA network. This is done by modifying
the network interfaces of the bot such that \texttt{eth1} has the same IP and
MAC addresses of the impersonated RTU, see center-left \tcircle{2} in
Figure~\ref{fig:attack-sub1}. Those steps are performed using a combination
of \texttt{ifconfig}, \texttt{brctl} and \texttt{route} Linux commands.
Then the bot starts a Modbus/TCP
server listening on the same IP and port of the real RTU server. The updating
period of the bot server is configurable via the time manipulation ($T_M$)
parameter. The server is implemented using
\texttt{pymodbus} and it runs in asynchronous mode using multiple threads. Once the
server is running then the Pub, Sub, and Forger kick in by running the
\texttt{mn.py} script. The script activates an asynchronous \texttt{paho} MQTT
client that let the bot joint the \Botnet network and enables packet spoofing on
the bot. Note that after these steps the bot is still connected to the RTU.

Let's assume that Figure~\ref{fig:attack-sub1} represents an impersonation
attack on the first substation. Let's see how a bot is using the
\CC Controller to send a valid Modbus/TCP \texttt{response} after a valid
\texttt{request} from the remote SCADA, see center-right \tcircle{2} in
Figure~\ref{fig:attack-sub1}. The bot intercepts the Modbus request
(Infiltrator), extracts the addresses of the requested values (Forger),
and passes them to the Pub component. The Pub publishes those addresses into the
relevant \texttt{sub1/rtu1} topics. The \CC Broker
module, which is orchestrating the \Botnet network, collects those messages. The \CC
Sub module then receives the messages containing the SCADA request addresses
from the Broker and pass them to the Controller module. The Controller
produces valid response values according to some estimation technique. A
discussion about some potential estimation techniques is presented in
Section~\ref{sec:learning}. The estimated response values produced by the
Controller are then
published by the \CC Pub module to the relevant \texttt{sub1/rtu1} topics. The
bot Sub module receives the estimated responses from the
\CC Broker and passes them to the Forger. Then the Forger creates
a valid Modbus/TCP response packet and sends it to the SCADA via the
Infiltrator.

The same impersonation technique is applied in each substation at
the same time. In general, this attack is \emph{distributed} because it
disconnects all the RTUs. It is \emph{coordinated} since each bot publishes
information about its controlled substation and subscribes to information
about other substations and about the status of other \Botnet devices. It is
\emph{cyber-physical} because once the
botmaster is able to impersonate all the RTUs then she can target the real
ones causing high-impact damages on the substations, while fooling the SCADA
server, see bottom-left \tcircle{3} in Figure~\ref{fig:attack-sub1}.


\subsection{Distributed Replay Phase}
\label{sec:replay}

The distributed replay attack is similar to a wireless
wormhole attack~\cite{hu06wormhole}. This type of attack enables a bot
to replay locally requests and responses that are coming from other remote
substations.

We now explain how to replay the content of a Modbus response
from the second substation to the first one. Let's assume that the target
request-response concerns the 100th holding register of the RTU in the second
substation. We assume
that the bots have already completed the initialization phase in the first
and second substation. Then,
the bot in the second substation will use a combination of \texttt{iptables}
and \texttt{libnetfilter\_queue} commands to extract in real time the hr100
payload from each valid RTU response. It will then
publish those payloads in the \texttt{sub1/rtu1/hrs/hr100} response topic.
The bot in the
first substation is subscribed to all the topics concerned sub1. Indeed it
will be able to reply to a hr100 SCADA request from the first RTU, with whatever
hr100 value is contained in the second RTU\@.

We note that the same technique can be used on different device's types, on
arbitrary sensor and actuator values and different substations at the same
time. The result is a coordinated cyber-physical attack that can potentially
alter the state of several substations without requiring detailed knowledge
of the physical process by the botmaster (\eg if attack is successful in one
substation replay it on the others).


\subsection{Evaluation of the \Botnet Attacks}
\label{sec:evaluation}


We performed a series of measurements on the \CC and two bots while conducting
the distributed impersonation and distributed replay attacks both in
the simulated and real \wadilong. For the network analysis, we used
Wireshark's built-in statistics and expert information to measure delays and
to flag anomalies in TCP connections. While we used Wireshark for convenience,
we note that detection rules in popular IDS, such as Bro and Snort will
work similarly~\cite{sanders2017practical}, so we expect the results to be
representative.

The main differences between the simulated and the real \Botnet
attacks are in terms of hardware. In the simulated attack, we used
MiniCPS~\cite{antonioli2015minicps} to simulate a \Botnet-based attack on
\wadilong. MiniCPS is a toolkit to perform real-time
cyber-physical system simulations using lightweight virtualization and it is based
on mininet~\cite{team2012mininet}. For the real attacks performed on the
\wadilong testbed (see Section~\ref{sec:wadi}), we used a commercial laptop running a
Linux OS to host the \CC station, and we modified Linux-based gateway
devices to act as bots. The \emph{same} code was run for the simulated and
real initialization, distributed impersonation and distributed replay phases.
What changed were the IP addresses (because of DHCP) and
the network interface names (since they are set by the OS).

Table~\ref{tab:attack-eval} lists the results of our evaluation using the
metrics presented in Section~\ref{sec:metrics}. The SCADA polling period is in
the order of few seconds ($T_S$) and we set the adversarial estimation period
($T_C$) and the maximum traffic manipulation periods ($T_M$) to approximately
be the half of it. Both simulated and real attacks generated one and four warning
messages about TCP packets with the reset flag set. We note that those few
warning messages could be avoided by improving the way the bot handles
existing TCP connections after the attack starts.

Recall that $\mu_d$ denotes the average delay introduced by the \Botnet.
In our experiments, $\mu_d$ was computed from the central SCADA server
comparing the response times while the system was and was not under attack.
Interestingly, $\mu_d$ \emph{resulted to be close to 0~ms}. This means that our centralized
\Botnet implementation did not cause significant delays in the real and
simulated SCADA system. We expect that this is due to our asynchronous communications
(\eg bots do not have to periodically wait for the messages coming from the
\CC) and our custom traits for the bots and the \CC (\eg bots implementation
is focused on the network manipulation while the \CC focuses on the
adversarial control).

A minor issue that we experienced attacking the real testbed is related
to $\Delta_{s}$ and $\Delta_{r}$ that are respectively the average time
difference between the last valid RTU packet and the first packet spoofed by
a bot, and the average time difference between the last SCADA request and
the first valid spoofed response by a bot. We have a significant discrepancy
between the attacks in the simulation framework (few milliseconds) and in the
real testbed (seconds). However, this situation is experienced only one time
when the attack is started, nonetheless, we are planning to conduct more
experiments to better investigate it. Finally, we note that the average CPU
load and memory (RAM) consumption on each bot is under 30\%. This should
allow performing the same attacks using devices that are even less powerful
than our infected gateway devices.

\begin{table*}[tb]
    \centering
    \caption{Evaluation of \Botnet attacks in a simulated environment and real testbed.
    $T_S$, $T_C$, $T_M$ are the SCADA, adversarial estimation and traffic
    manipulation periods.
    $\Delta_{s}$ is the average time difference between the last valid RTU packet and the first
    packet spoofed by a bot.
    $\Delta_{r}$ is the average time difference between the last SCADA request and the first
    valid spoofed bot response.
    $\mu_{d}$ is the additional average delay introduced by the \Botnet measured
    from the SCADA server.
    $n_{e}$ is the number of Wireshark's expert info warnings and error
    messages.
    $\mu_{CPU}$ and $\mu_{RAM}$ are the approximate average CPU and RAM load
    on each bot. $n_B$ is the number of controlled substations.
    Our optimized \Botnet implementation is able to attack the system with delay ($\mu_d = 0$) as measured at the SCADA server.}
    \label{tab:attack-eval}
    \begin{tabular}{lrrrrrrrrrrr}
    \toprule
      \textbf{Attack} & $T_S$ [s]& $T_C$ [s] & $T_M$ [s] & $\Delta_{s}$ [ms] &
        $\Delta_{r}$ [ms]& $\mu_{d}$ [ms] & $n_{e}$ &
        $\mu_{CPU}$ & $\mu_{RAM}$ & $n_B$ \\
    \midrule
    Simulated Distributed Impersonation & 1.5 & 0.8  & 0.6  & 2.6 & 3.6 & 0.0 & 1 &
        2\% & 20\% & 2    \\
    \midrule
    Real Distributed Impersonation & 1.0 & 0.5 & 0.5 & 7000 & 7010 & 0.0 & 4 & 10\% & 30\%
        & 2\\
    \midrule
    Simulated Distributed Replay & 1.5 & 0.8  & 0.6  & 43 & 44 & 0.0 & 1 &
        2\% & 20\% & 2    \\
    \midrule
    Real Distributed Replay & 1.0 & 0.5 & 0.5 & 7071 & 7069 & 0.0 & 4 & 10\% & 30\%
        & 2  \\

    \bottomrule

    \end{tabular}
\end{table*}



\section{Discussion}
\label{sec:discussion}

In this section, we discuss different ways to develop an
adversarial control strategy, several methods to increase the
stealthiness of our \Botnets and some techniques to optimize
the presented attack.

\subsection{\Botnet: Control Strategies}
\label{sec:learning}

So far, we have not discussed how the attacker should develop a
suitable control strategy to take over the physical process state. We
now briefly outline different potential approaches. Implementing and
evaluating those approaches is out of the scope of this work, and we plan
to do so in future work.


In general, the attacker has to learn how the physical process reacts to
changes in actuator states, and how the state of the physical process
is transitioning over time. Such knowledge can be obtained through
physical process estimations~\cite{garcia2017use,ljung1998system}, manual
data analysis, and machine learning approaches (e.g., similar to the
ones employed for attack detection for CPS~\cite{ozay2016machine}).
General learning of target CPS infrastructure has been discussed
in~\cite{feng16characterizing}.

Based on a solid understanding of the process, the attacker needs to find
a sequence of actions that will lead to the goal state of the system,
potentially while considering legitimate control reactions that are not under
the influence of the attacker. In addition, it is likely that the physical
process simulation would be optimized to remain hidden from process observers
for as long as possible, or reaches its goal as soon as possible (related
tradeoffs are discussed in~\cite{urbina2016limiting}). Commonly, such control
strategies require continuous tracking of the process physical evolutions
using state estimation techniques such as Kalman filters and Luenberger
observers.







\subsection{\Botnet: Stealthiness}
\label{sec:stealthiness}


There are several ways to increase \Botnet stealthiness. We present a brief
discussion about two strategies:

\Paragraph{Stepping stones} We might want to introduce extra devices
as intermediate proxies in the communication between the bots and the
\CC~\cite{khattak2014taxonomy} over the Internet. Those extra nodes might
 introduce secure tunnels (\eg use Tor~\cite{sanatinia2015onionbots}). With this
solution, we pay a penalty in terms of botnet latency and we should take into
consideration if the tradeoff is worth.

\Paragraph{\CC Protocol} We understand that using a protocol for the
\CC channel that is different from the target one might increase the
possibility of detection. However, this is the case also when the \CC is using
the same protocol as the target system~\cite{vormayr2017botnet}. Two solutions to
mitigate this problem are encryption (\eg TLS) and obfuscation (\eg
obfuscated data structures for MQTT messages).

\section{Related Work}
\label{sec:related}

We've seen novel designs of botnets from the cyber-security field. For
example, DNS botnet~\cite{anagnostopoulos2013dns}, structured and unstructured
peer-to-peer botnets~\cite{wang2010advanced,rossow2013sok}, server-less
botnets~\cite{zhao2017serverless}, botnet-as-a-service~\cite{chang2014baas},
mobile botnets~\cite{mtibaa2015mobibot}, bitcoin-powered
botnets~\cite{ali2015zombie} and botnets pivoting from social
networks~\cite{compagno2015elisa}. However, those designs are not addressing
cyber-physical systems and OT networks.

There are several interesting analysis of traditional IT botnets.
In~\cite{stone2011analysis}, the authors managed to act as fake \CC servers and
collected information about the Torpig botnet. In~\cite{abu2006multifaceted},
the authors presented a system able to capture and track more than 100 unique
IRC-based botnets to measure the percentage of malicious traffic attributed
to those botnets on the Internet. In~\cite{nadji2013beheading}, the authors
proposed a botnet take-down analysis and recommendation system. However, none of
those papers analyzes a cyber-physical botnet with suitable quantitative
metrics such as latency and size of the \CC packets.

There are recent academic works about cyber\-/physical attacks targeting several CPS
devices. In particular, preferred targets are programmable logic controller
(PLC). Authors discussed ran\-som\-wa\-re~\cite{formby17ransomics}, firmware
modifications~\cite{cui2013firmware}, rootkits~\cite{abbasi2016ghost},
physics\-/aware malware~\cite{garcia2017hey}, and stealthy
Man\-/in\-/the\-/Middle~\cite{urbina2016attacking} attacks. Recently, we
have seen in the wild targeted attacks on Safety Instrumented Systems
(SIS)~\cite{dragos2017trisis}.
Those attacks target a single device and they are not performed using
(cyber-physical) botnets.

We have seen also attempts to detect botnets for CPS. In particular,
in~\cite{yang2017scada} the authors are trying to detect P2P SCADA botnets by
means of custom network monitoring. However, they assume to be attacked by a
traditional P2P botnet.

\section{Conclusions}
\label{sec:conclusions}


In this work, we argue that adversarial control attacks on CPS requires a
novel class of botnets that we define as \emph{cyber-physical}. Those botnets
have different requirements from traditional IT botnets such as usage of
adversarial control strategies, coordinated interactions among the bots, and
additional constraints from the target system. 



To address those challenges we presented \Botnet: a framework to build
cyber-physical botnets. We leverage on a \emph{publisher-subscriber} paradigm
for the \CC channel to coordinate our bots with minimal overhead. We define
an orthogonal set of \emph{traits} to customize our bots and \CCs according
to their role in the attack and their hardware and software capabilities. For
example, the bots might be specialized for packet manipulation while a \CC
focuses on generating adversarial control decisions. \Botnet allows using
different adversarial control strategies like machine learning classification,
real-time simulation, and Kalman filtering estimation. Furthermore, we are
able to adapt our botnets to different network architectures, protocols, and
physical processes.

We showed the design of a centralized botnet to attack a centrally controlled
CPS and a decentralized botnet to attack a CPS with distributed control. We
implemented the former using MQTT for the \CC protocol and Modbus/TCP as
the target network protocol. We evaluate our implementation by performing
two coordinated cyber-physical attacks: \emph{distributed eavesdropping},
and \emph{distributed impersonation}. We evaluate our attacks with custom
cyber-physical botnets metrics and we showed that our \Botnet introduces zero
additional delay ($\mu_d = 0$) while the system is under attack. As result,
\Botnet is able to conduct attacks that cause minimal temporal changes to the
traffic, which hide the manipulation from operational alarms that might be in
place.


We expect those findings on capabilities will raise awareness with
stakeholders of threatened systems, and allow the defenders to design
more suitable countermeasures. Potential countermeasures against our
\Botnet would include close monitoring of physical process states with
hardened sensors, general hardening of industrial devices against
exploitation, and network segmentation and monitoring.



\bibliographystyle{plain}
\bibliography{bibliography}
\balance


















\end{document}